\documentclass[prd,twocolumn,superscriptaddress,altaffilletter,amssymb,showpacs,nofootinbib]{revtex4}
\usepackage[dvips]{graphicx}
\usepackage{amsmath}
\newcommand{\be}{\begin{equation}}
\newcommand{\ee}{\end{equation}}
\newcommand{\bea}{\begin{eqnarray}}
\newcommand{\eea}{\end{eqnarray}}

\begin{document}

\title{Causality and Unitarity Are Not Violated in Space-Times with an Additional Compact Time-like Dimension}

\author{Israel Quiros}\email{israel@uclv.edu.cu}
\affiliation{Universidad Central de Las Villas, Santa Clara CP 54830, Cuba}

\date{\today}
\begin{abstract}
The hypothesis that the causal properties of space-time, as well as other properties of physical systems like unitarity, charge conservation, etc., might be decided by the higher dimensional structure (in particular, higher-dimensional physical laws), rather than by the four-dimensional one, is explored in order to evade the most awkward problems of higher-dimensional theories with compact extra time-like dimensions: violation of causality and of unitarity. 
\end{abstract}

\pacs{03.50.-z, 04.20.Cv, 04.50.+h, 11.25.Mj, 98.80.-k, 98.80.Cq}

\maketitle

\section{Introduction}

Unified theories of the fundamental interactions require the existence of compact extra dimensions beyond the four space-time dimensions that are experienced by the standard model particles (SMP). It is a widespread belief that these compact extra dimensions have to be space-like. The belief might be related with the awkward problems associated with violations of causality and of unitarity, that are inherent in theories with compact time-like extra dimensions \cite{yndurain}. Notwithstanding, one can find some scattered papers where time-like additional dimensions are used either to find a cure to the cosmological constant problem \cite{arefeva}, or to study strings \cite{popov,bars1} and supersymmetry \cite{bars2}.

Time-like extra dimensions have been used within the framework of brane world models, as an alternative to reconcile the mass hierarchy problem resolution with correct cosmological expansion of the visible universe within Randall-Sundrum type of brane world \cite{chaichian}, and, more generaly, to show that neither the cosmological big-bang singularity, nor big-crunch arise in brane worlds with an extra time-like dimension \cite{sahni}. Time-like extra dimensions have been invoked, also, to generalize the standard model of particles and forces \cite{bars4}.\footnotemark\footnotetext{To our knowledge the most systematic approach to extra time-like dimensions have been undertaken by I Bars and co-workers who have developed the so called notion of 2T physics \cite{bars1,bars2,bars3,bars4}.} 

There are phenomenological bounds imposed by experiments to the feasibility of time-like extra dimensions \cite{yndurain,dvali,erdem,matsuda}.\footnotemark\footnotetext{These bounds are not applicable to the 3-brane solitonic model developed in reference \cite{iglesias}, where there are no propagating tachyons or negative norm states in the bulk.} These bounds depend on the topology of the extra (time) space and are set by violations of causality and unitarity. In the simplest case when there is just an additional time-like dimension of circular topology (as in standard Kaluza-Klein mechanism), the argument is the following \cite{yndurain}: an extra time-like dimension curled with a radius $L$, produces modifications in the particle propagators. For example, for the propagator of a photon one has:

\be D_{\mu\nu}(k,l)=-i g_{\mu\nu}\frac{1}{k^2+l^2+i\epsilon},\label{propagator}\ee where $k^2=k_0^2-{\bf k}^2$ are the four-momenta conjugated to the ordinary space-time coordinates $x^\alpha=(x^0,{\bf x})$, while $l=n/L$ ($n=0,\pm 1,\pm 2,...$) is the momentum conjugated to the extra time-like dimension. In addition to the photon ($n=0$), one has a tower of excitations of imaginary masses quantized in units of $i/L$. These tachyonic modes are the ones that originate causality violations in the theory. Violation of unitarity arises because, in addition to the ordinary photon pole at $n=0$, one obtains unphysical poles at $k^2=-n^2/L^2-i\epsilon$ ($n=\pm 1,\pm 2,...$). These modify the Coulomb potential in such a way that it becomes complex, causing violation of probability \cite{yndurain,matsuda}. The argument can be straightforwardly extended to an arbitrary number of extra time-like dimensions. 

When non-compact extra time-like dimensions are being considered (for instance, in brane models where the SMP are localized in extra time), the argument is similar. Although, in this case, the KK modes of the SMP are not present, the KK gravitons propagating in the bulk are viewed by us as tachyonic states (see, for instance, reference \cite{dvali}).

In the present letter we want to approach the subject from a completely different perspective. Our approach relies on a very simple (perhaps naive) hypothesis: 

\begin{itemize}
	\item The causal properties of space-time and other properties of (quantum) physical systems such as unitarity, charge and energy conservation, etc., are decided by the higher-dimensional structure rather than by the four-dimensional one.
\end{itemize} Exploring the relevance of this hypothesis to the feasibility of time-like extra dimensions, is the main objective of the present note. The implications of the hypothesis for other aspects of particles physics and gravity will be also briefly commented on. 

\section{A comment and The set up}

We want first to make a comment on our hypothesis. Usually causality is analysed from the point of view of four-dimensional (4D) observers like us. However, if extra dimensions are taken into account one has to wonder whether the causality (and other issues) is decided by 4D observers or by the higher-dimensional structure. Consider, for instance, the case of brane worlds where the SMP are confined to a 3-brane while the graviton may propagate in the extra dimensions (for definiteness consider space-like extra dimensions). In this case the question is trivial: the higher-dimensional structure decides causality. In fact, although 4D (brane) observers may see causality violations, these might not be actually violations of causality. Instead, provided the brane is appropriately bent, these may be explained as due to propagation of the graviton from a point on the brane into a different point (also on the brane), through a shorcut in the bulk (higher-dimensional) space-time. However, when compact extra time-like dimensions are considered, the discussion about causality is not that clear, and we have to rely on some hypothesis given a priori (like the one we are exploring in this letter).

Here, for simplicity, we shall be concerned with a 5-dimensional (5D) space-time with two time coordinates $x^0=t$, and $x^5=\sigma$ respectively. The extra-time coordinate $\sigma$ will be assumed to have circular topology ($L$ being the radius of the circle). As a consequence the fields will be periodic in $\sigma$. Unless the contrary is specified, we shall neglect the effects of gravity. In correspondence, the following (flat) metric tensor will be considered: $\eta_{AB}=(\eta_{00}=-1,\eta_{ik}=\delta_{ik},\eta_{55}=-1)$. Capital Latin indexes run through 5D space-time $A,\;B,...=0,1,2,3,5$, while small indexes run through three-dimensional space $i,\;k,...=1,2,3$, and Greek indexes run through ordinary 4D space-time $\alpha,\;\beta,...=0,1,2,3$. The line element of the 5D space-time can be written in the following form:

\be ds_5^2=-dt^2+\delta_{ik}dx^idx^k-d\sigma^2=ds^2-d\sigma^2,\label{linelement}\ee where $ds^2=\eta_{\mu\nu}dx^\mu dx^\nu=-dt^2+\delta_{ik}dx^idx^k$ is the ordinary 4D line element experienced by four-observers like us. 

Notice that, since the extra space is spanned by a time coordinate, the motion of any higher-dimensional as well as of any 4D object, is confined to the Euclidean three-dimensional space with metric $\delta_{ik}$. The meaning of this is that we do not need of any confining mechanism to constrain the motion of (for instance) standard model particles to our physical three-space. Even gravitation (or other possible higher-dimensional fields) is confined to the common to all particles three-dimensional Euclidean space which we live in. Actually, as it will be discussed in more detail later on in this letter, when small perturbations of the metric $\eta_{AB}$ are considered, the static weak field limit of the higher-dimensional Einstein's equations (meaning, in particular, that field variables do not depend neither on $t$ nor on $\sigma$) leads to a standard 3-dimensional Poisson equation, whose solution is the customary gravitational potential $V(r)\propto 1/r$. This point is usually missing when extra time-like dimensions are being considered.

\section{Causality and Unitarity}

The causal structure of the 5D space-time is defined by the line element (\ref{linelement}), so that 5D causal propagation is subjected to the condition that $ds_5^2\leq 0$. In correspondence with this, the 5D photons will propagate with the maximal speed $ds_5^2=0\;\Rightarrow\;c_5^2=\delta_{ik}(dx^i/dT)(dx^k/dT)=1$ ($dT=\sqrt{dt^2+d\sigma^2}$). The speed of a 5D photon as measured by a 4D observer (like us), can be computed by considering that $dT=dt\sqrt{1+\xi^2}$ ($\xi\equiv d\sigma/dt$):

\be (c_{obs})^2=\delta_{ik}\frac{dx^i}{dt}\frac{dx^k}{dt}=c_5^2(1+\xi^2)=1+\xi^2,\label{lightspeed}\ee where $c_{obs}$ is the speed of the 5D phton as measured by a four-observer and the parameter $\xi$ is a measure of five-dimensionality. Notice that, although a 5D photon always moves at the maximal speed $c_5=1$, it is seen by a 4D observer as moving at a superluminal speed ($c_{obs}>1$) so that, in consequence, it is assumed to be a tachyon that is responsible for undesirable violations of causality. The question is whether these violations are just a consequence of the unseen extra time coordinate and are not dangerous for the behaviour of physical systems, or, instead, are actual violations and have (negative) impact on the 4D physics. 

It is straightforward realising that the answer to the above question depends on one's choice of the basic structure: if the basic structure defining the behaviour of physical systems is the higher-dimensional one (as it is the case in the present note thanks to our hypothesis), the answer is that the aforementioned violations are artifacts of the higher-dimensional structure and have no impact on the physics. 

It will be very instructive to investigate the following oversimplified situation. Let us consider a free propagating 5D (massless) photon that obeys 5D laws that are a minimal extension of Maxwell's theory: 

\be \Box_{(5)}A_A\equiv\eta^{NM}\partial_N\partial_M A_A=0,\label{5dphotoneq}\ee where $A_A$ is the vector potential of the 5D electromagnetic field. The above 5D law can be further written in a form that is manifestly four dimensional and, therefore, is adequated to a 4D observer (like us). For this purpose we consider the following "Kaluza-Klein" (KK) decomposition that is dictated by the topology of the extra-time dimension ($n=0,\pm1,\pm2,...$):

\be A_A(x,\sigma)=\sum_{n} A^n_A(x)\;e^{in\sigma/L}.\label{decomposition}\ee We obtain

\bea &&\bar\Box A^n_\mu(x)+(n/L)^2 A^n_\mu(x)=0,\label{4dphotoneq}\\
&&\bar\Box\phi_n(x)+(n/L)^2 \phi_n(x)=0,\label{4dscalarphoton}\eea where $\bar\Box\equiv\eta^{\mu\nu}\partial_\mu\partial_\nu$ is the standard (flat) 4D D'Lambertian, and $\phi_n(x)\equiv A^n_5(x)$. As seen, the zeroth mode ($n=0$) corresponds to the ordinary 4D (massless) photon, while massive modes with negative massess squared $M^2_n=-n^2/L^2\;\Rightarrow\;M_n=in/L$, appear to a 4D observer as tachyonic degrees of freedom moving with speeds greather than that of light $c=1$. These are the responsibles for the above mentioned violation of causality. 

In respect to a 4D observer unitarity is also violated. Actually, consider for instance, the scalar modes $\phi_n(x)$. In the rest frame the amplitude of these modes can be written as (recall that a standard 4D observer is not aware of the existence of the extra-time): 

\be \phi_n(x)=\phi_{n0} e^{iM_n t}=\phi_{n0} e^{-n t/L}.\label{4damplitude}\ee The norm of these states can be written as:

\be |\phi_n|^2=|\phi_{n0}|^2\;e^{-2n t/L},\label{4dnorm}\ee so, in a time $\tau_n\sim L/n$ the corresponding mode decays, apparently into "nothingness", causing that probability is not conserved. From the phenomenological point of view the 4D observer has to assume $\tau_1$ to be of the order of the Plack time ($\tau_1\sim 0.1 M^{-1}_p\approx 10^{-20} MeV^{-1}$),\footnotemark\footnotetext{This constrain is necessary in order to be consistent with the phenomenological bound imposed by the consideration of baryon stability in the nuclei \cite{yndurain}.} so that violations of causality and unitarity due to the presence of the tachyonic states are not observed and have no major impact on the 4D physics. Higher order KK states with $|n|>1$, are more unstable and have less influence on 4D observations.

However, according to our hypothesis, the physics is dictated by the higher-dimensional (5D) structure, so that, a 5D observer is able to notice the missinterpretation of the physical reality by the 4D observer. In fact, to a five-dimensional observer, who is aware of the existence of the extra-time $\sigma$, in the rest frame (see the decomposition (\ref{decomposition})):

\be \phi_n(x)=\phi_{n0} e^{in\sigma/L},\label{2tamplitude}\ee so that the norm of the corresponding states do not depend on time (compare with (\ref{4dnorm})):

\be |\phi_n|^2=|\phi_{n0}|^2,\label{2tnorm}\ee i.e., probability is conserved (unitarity is not violated). 

Another way to explain this situation is the following. Consider, for simplicity, a 5D scalar field $\psi(x^A)$ of mass $m$. As before, let us assume the KK decomposition that is dictated by the topology of the extra-time dimension: $\psi(x,\sigma)=\sum \psi_n(x)\exp{(in\sigma/L)}$. The corresponding 5D KG equation $(\Box_{(5)}-m^2)\psi(x,\sigma)$ can be written in manifest 4D form:

\be (\bar\Box-m_n^2)\psi_n(x)=0,\label{scalarfieldkg}\ee where we have accommodated the $(n/L)^2$ term- coming from the derivatives in respect to the extra-time $\sigma$- in the squared mass term $m_n^2\equiv m^2-(n/L)^2$. The indeterminacy in the sign of this term is the origin of possible causality and unitarity violations. Notice, however, that since the extra space is time-like, it is perhaps more appropriate to group the term $(n/L)^2$ under energy squared instead. To see this point, consider the plane wave approach:

\be \psi_n(t,\textbf{x})= \psi^0_n e^{-i(Et-\textbf{px})},\label{planewave}\ee where $E$ is the energy, and $\textbf{p}$ the momentum of the field. Then, equation (\ref{scalarfieldkg}) can be written, alternatively, in the following way:

\be E_n^2-|\textbf{p}|^2-m^2=0,\ee where now the energy of the $n$- mode of the field is given by $E_n^2=E^2+n^2/L^2$. Notice that under this alternative 4D interpretation there are neither causality nor unitarity violations coming from the extra-time. Instead, the energy of each individual mode of the field gets increased by the factor $\sqrt{1+(n/EL)^2}$.\footnotemark\footnotetext{Perhaps a more compact and clear way to explain the above picture is by considering particle propagators. Actually, since according to our hypothesis 5D laws are to be obeyed by (quantum) physical systems, then, assuming minimal extension of the known laws, the 4D photon propagator (\ref{propagator}) has to be replaced by the following 5D propagator: $D_{AB}=-i\eta_{AB}\frac{1}{k_{(5)}^2+i\epsilon}$, where $k_{(5)}^2=k^2+l^2$- the photon's 5D momentum, is the observed quantity instead of $k^2=k_0^2-{\bf k}^2$. In consequence no unphysical poles that may lead to modifications of the Coulomb potential arise. Hence, contrary to the argument of \cite{yndurain} (see also in \cite{erdem}), there is no associated nucleon decay mode, i.e, the decay width vanishes.}

In consequence, if our hypothesis is correct, previously considered phenomenological bounds on the feasibility of additional (compact) time-like dimensions \cite{yndurain,dvali,erdem,matsuda} are incorrect.

\section{Newtonian gravity}

Consider small perturbations of 5D flat metric: $\eta_{AB}\rightarrow \eta_{AB}+h_{AB}(x^C)$. Assuming minimal extension of general relativity, in the transverse, traceless gauge ($\partial^Nh_{NA}=h=0,\;\;h\equiv\eta^{NM}h_{NM}$), the Einstein's field equations read:

\be \Box_{(5)}h_{AB}=8\pi T_{AB},\label{5dgraviton}\ee where $T_{AB}$ is the stress-energy tensor of other matter degrees of freedom than the graviton. Recalling that the Newtonian gravitational potential $V=h_{00}/2$, and considering the static limit, i.e., there is no dependence of the field variables neither on $t$ nor on $\sigma$, equation (\ref{5dgraviton}) coincides with the standard Poisson equation for the Newtonian potential:

\be \nabla^2 V=4\pi\rho,\label{poisson}\ee where $\rho$ is the energy density of the matter degrees of freedom. The later equation yields to the customary expression for the Newtonian potential surrounding a given matter distribution with spherical symmetry: $V(r)=4\pi/|\textbf{r}|$. In consequence, in the weak, static field limit, Newton's law of gravity is not modified in any way by the presence of the extra-time $\sigma$. This means that we have not to rely neither on Kaluza-Klein compactification, nor on brane concept or other alternative mechanisms to recover 4D Newtonian gravity. 

An important consequence that emerges from the above analysis is that phenomenological bounds imposed on the feasibility of time-like extra-dimensions that have been considered so far, are correct only if causal behaviour (and other properties of physical systems) is dictated by the 4D structure. If the higher-dimensional structure is the one that decides the dynamics, instead, these phenomenological bounds are fictitious, and, correspondingly, there could be fictitious violations of causality and unitarity seen by 4D observers that are, actually, artifacts of the higher-dimensional structure. Only a class of observers that are aware of the existence of the time-like extra dimensions (in particular 5D observers), can understand the whole physical picture.

\section{Vacuum energy and phenomenology of an extra-time dimension}

It is important to stress, that the picture resulting from our hypothesis is different from the standard Kaluza-Klein mechanism in several aspects. In the first place, we apply the KK decomposition (a neccessary consequence of the asumed topology of the extra-time dimension) not to the effective four-dimensional fields, but to the higher-dimensional fields instead (see, for instance, the decomposition (\ref{decomposition})). In the second place, the dynamics of the fields is governed by 5D laws (equations like, for instance, (\ref{5dphotoneq})), instead of effective 4D equations obtained after applying the decomposition (see (\ref{4dphotoneq},\ref{4dscalarphoton})). In the third place, since no known stringent phenomenological bound needs to be imposed, the size of the extra time-like dimension does not have to be necessarily small. Therefore, the KK compactification procedure does not have to be necessarily applied. This is, precisely, the clue to impose some phenomenological bounds to the possible effects of an additional time-like dimension by computing the contribution of zero-point fluctuations of the fields to the energy density of empty space. This will be our next goal.

Although, in principle, to compute the contribution of zero-point fluctuations to the vacuum energy one has to consider the quantum theory of 5D fields (a subject that is much behind the scope of the present letter), a crude estimate can be given by replacing the standard equation for vacuum energy density:

\be \left\langle\rho_v\right\rangle=\frac{1}{(2\pi)^3}\int_0^\Lambda d^3p\frac{1}{2}\sqrt{p^2+m^2},\label{standard}\ee where $\Lambda$ is some momentum cutoff, usually assumed to be of the order of the Planck mass, by the following modified equation:

\bea &&\left\langle\rho_v\right\rangle=\sum_n \left\langle\rho_v\right\rangle_n,\nonumber\\
&&\left\langle\rho_v\right\rangle_n=\frac{1}{(2\pi)^3}\int_0^\Lambda d^3p\frac{1}{2}\sqrt{p^2+m^2-n^2/L^2}.\label{modified}\eea Next we turn to the continuous limit $L\rightarrow\infty$, meaning that we drop compactness of the extra-time dimension and assume this additional direction to be of very large (perhaps infinitely large) extent, as discussed in the beginning of this section. In consequence, the symbol $\sum$ may be replaced by $L \int dy$, where the new variable $y=n/L$ has been considered. In general one has

\bea &&\sum_n \left\langle\rho_v\right\rangle_n\rightarrow L\int_{-\infty}^\infty dy \left\langle\rho_v(y)\right\rangle,\nonumber\\
&&\left\langle\rho_v(y)\right\rangle=\frac{1}{(2\pi)^3}\int_0^\Lambda d^3p\frac{1}{2}\sqrt{p^2+m^2-y^2}.\eea Since the expression under the integral in the later equation is an even fucntion of $y$, then, one may replace $\int_{-\infty}^\infty\rightarrow 2\int_0^\infty$. Besides, as customary, one has to set some "momentum" cutoff $\lambda$ so that $\int_0^\infty\rightarrow\int_0^\lambda$. We obtain (compare with equation (\ref{standard})):

\be \left\langle\rho_v\right\rangle=\frac{2L}{(2\pi)^3}\int_0^\Lambda d^3p \int_0^\lambda dy \frac{1}{2}\sqrt{p^2+m^2-y^2}.\label{contivac}\ee Notice there are now two energy scales $\Lambda$ and $\lambda$. Taking the integral in $y$, and considering the limit $\lambda\ll\sqrt{p^2+m^2},\;\Lambda\gg m$, one gets:

\be \left\langle\rho_v\right\rangle\approx\frac{L\lambda}{(2\pi)^3}\int_0^\Lambda d^3p \sqrt{p^2+m^2}\approx\frac{L\lambda\Lambda^4}{8\pi^2}.\label{computation}\ee 

If we compare equation (\ref{computation}) with equation (3.5) of reference \cite{weinberg}, where the factor $2L\lambda$ is missing, we can see that the usual hughe value of the vacumm energy density obtained by using (\ref{standard}), can be made as small as desired by apropriately giving the value of the product $L\lambda$. Actually, assuming $\Lambda^4\approx 10^{73}GeV^4$ ($\Lambda$ is nearly the Plack mass), one can easily check that the almost vanishing (but non-null) observed value of the energy density of empty space: $\left\langle\rho_{vac}\right\rangle\approx 10^{-47}GeV^4$ can be obtained by setting $L\lambda\sim 10^{-118}$. Therefore, correct phenomenology of vacuum energy requires that the energy scale $\lambda$ be almost vanishing, but non-null. An exact four-dimensional world, meaning that $\lambda=0$, would entail an exactly vanishing vacuum energy. 

The fact that the observed value of the vacuum energy (cosmological constant) is not exactly null, means that the extra time-like dimension leaves its signature in the observed 4D world where we live. On the other hand, to all practical purposes, the space-time we live in is four-dimensional since $\lambda^2=0$ with more than enough accuracy.

\section{Concluding remarks}

In the present letter we have investigated causality and unitarity in space-times with an extra time-like (compact) dimension from a different perspective than in former approaches. Our approach relies, mainly, on three assumptions: i) The space-time is five-dimensional (two time-like dimensions plus the ordinary three spatial dimensions), ii) the following hypothesis holds true: the main properties of (quantum) physical systems are decided by the higher-dimensional structure, and iii) the five-dimensional laws are a minimal extension of ordinary 4D laws.

If the former assumptions are correct, then, we can conclude that formerly discussed violations of causality and unitarity arising in space-times with an extra compact time-like dimension are fictitious. These are no more than artifacts of 4D description of physics, due to ignorance of the additional time direction. In particular none of the previously considered phenomenological bounds are applicable any more. This means that the extra time-like dimension does not have to be neccessarily small. The consequence is that we do not need of any KK compactification mechanism, or brane construct, or any other confining mechanism, to explain standard 4D behavior as seen by an ordinary four-dimensional observer.

Not very stringent phenomenological bounds are imposed by vacuum energy: In order to obtain the enormously small observed value of the vacuum energy (i.e., in order to avoid the cosmological constant problem \cite{weinberg}), the cutoff $\lambda$ on the extra time-like momentum, has to be incredibly small. This means that to all practicall purposes our world is a four dimensional space-time. 

It is apparent that, since the value of the vacuum energy had to be huge in early stages of the cosmic evolution, the existence of an extra time-like dimension could has important impact on the outcome of the cosmic hystory. This will be the subject of forthcoming work.

Our hypothesis, if correct, might have a major impact on the physics of standard model particles. It might be that exploring these issues would shed some light on the validity of the present approach. Perhaps the work along the lines of \cite{bars4} could be considered as a possible hint to study the consequence of our hypothesis for the standard model of the fundamental interactions. 

\acknowledgements The author thanks the MES of Cuba by partial financial support of the present research.

\end{document}